\documentclass[aps,prl,reprint,groupedaddress]{revtex4-1}
\usepackage{amsmath,amssymb,graphicx}

\begin{document}

\title{Phase diagram for injection locking a superradiant laser}


\author{Kevin C. Cox, Joshua M. Weiner, and James K. Thompson}
\affiliation{JILA, NIST, and University of Colorado, 440 UCB, 
Boulder, CO  80309, USA}


\newcommand{\Jarb}{{\bf J}(\theta,\phi,C)}
\newcommand{\Jperp}{{\bf J}_{\perp}}
\newcommand{\Jperpmag}{J_{\perp}}
\newcommand{\Jequator}{{\bf J}(\pi/2,\phi,C)}
\newcommand{\Jphionly}{{\bf J}(\pi/2,\phi,1)}
\newcommand{\bra}[1]{\ensuremath{\left\langle {#1} \right|}}
\newcommand{\ket}[1]{\ensuremath{\left|  #1 \right\rangle}}

\date{\today}

\begin{abstract}
\noindent We experimentally and theoretically study the response of a superradiant or bad-cavity laser to an applied coherent drive.  We observe two forms of synchronization (injection locking) between the superradiant ensemble and the applied drive:  one attractive and one repulsive in nature.  We explain the region of repulsion as arising from the higher three-dimensional description of the atomic spin state that stores the laser coherence in a superradiant laser, as opposed to a two-parameter description of the electric field in a traditional good-cavity laser.  We derive a phase diagram of predicted behavior and experimentally measure the response of the system across various trajectories therein.
\end{abstract}


\maketitle

\begin{figure}[h!]
\includegraphics{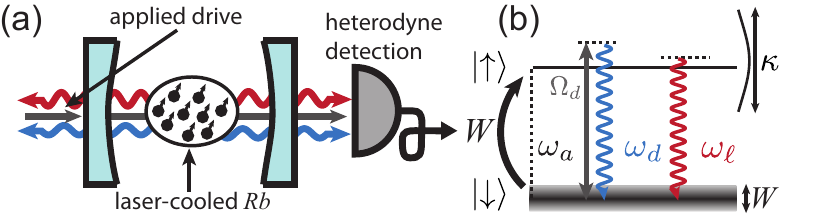}
\caption{\label{fig:fig1} Experimental setup and level diagram.  (a) Atoms interact with both the externally applied drive (grey) and the intra-cavity field generated by their collective emission (blue and red).  The superradiant laser primarily responds at two frequencies, the drive frequency $\omega_d$ and a self-lasing frequency $\omega_{\ell}$.  
(b)  The characteristic frequencies are displayed in a level diagram, and all lie within one cavity mode of width $\kappa$.  The Raman laser system is approximated as a 2-level laser incoherently repumped through intermediate optically excited states (not shown) at rate $W$.  $W$ is also the primary source of broadening of the lasing transition.  In this work, the ratio of atomic and optical linewidths is $W/\kappa \approx 5\times 10^{-2}$ to $5\times 10^{-3} \ll 1$, placing the system deep into the bad-cavity or superradiant regime.  The state $\ket{\uparrow}$ is a dressed state consisting of a ground hyperfine state of Rb coupled non-resonantly to an optically excited state as described in \cite{Bohnet2012,Bohnet2012a,Weiner2012,Bohnet2013}.  The applied drive couples $\ket{\downarrow}$ and $\ket{\uparrow}$ with an on-resonance Rabi frequency $\Omega_d$.}
\end{figure}

\noindent In a superradiant (or ``bad-cavity'') laser, the atomic coherence decays much more slowly than the optical cavity field.  As such, the atomic coherence primarily stores the laser's phase information and is initially established via spontaneous synchronization of the individual atomic dipoles (as in Fig. \ref{fig:fig1}(a)).  Unlike in conventional good-cavity lasers, coherence has been shown to persist with less than one, and even zero, intracavity photons \cite{Bohnet2012,Bohnet2013,Weiner2012}.  This bad-cavity regime of laser physics has generated recent interest because it offers a promising route for overcoming fundamental thermal mirror noise in order to realize laser linewidths of one milliHertz or less \cite{Meiser2009}.  

More broadly, cold atom-cavity systems are extremely well-controlled experiments useful for observing many-body phenomena with the cavity mode providing strong long-range interactions between the atoms.  For example, the spontaneous spatial ordering \cite{Inouye23071999,PhysRevLett.91.203001,PhysRevLett.98.053603,Greenberg:11} and realization of the Dicke model \cite{Esslinger2010} in cold atom-cavity systems are examples of nonequilibrium phase transitions and provide insights into our fundamental understanding of phase transitions in condensed matter physics \cite{PhysRevA.87.023831}. Further,  atomic ensembles coupled to many cavity modes may allow the creation of exotic phases of matter with emergent crystallization and frustration \cite{PhysRevLett.107.277201,PhysRevLett.107.277202}, and could serve as a model system for associative memories \cite{doi:10.1080/14786435.2011.637980}.  Superradiant lasers have been identified as an interesting system in which to study the problem of synchronization of quantum oscillators \cite{Xu2014,Manzano2013,PhysRevLett.112.094102}.  In all, superradiant atom-cavity systems and related systems promise continuing interest for both technological and fundamental reasons.

In this Letter, we study synchronization of a cold atom-based superradiant laser to an externally applied optical field that is injected into the lasing cavity mode (Fig. \ref{fig:fig1}(a)).  The synchronization is analogous to injection locking in a good-cavity laser, but in this superradiant system phase locking is manifested as collective synchronization of an ensemble of cold atoms to the applied drive.  For a weak applied drive, the laser inversion is approximately constant and the system can be approximately mapped to a driven Van der Pol self-oscillator, a canonical system in synchronization physics \cite{doi:10.1142/S0218127400001481, pikovsky2003synchronization}.  We directly observe the two synchronization behaviors predicted for such a system.  However, when the applied drive or detuning of the drive are large, we observe additional effects such as frequency repulsion of the self-lasing and ultimately the suppression of all stimulated power.  These two effects arise from the full three-dimensional description of the atomic spin or Bloch vector, in comparison to a two-dimensional description of a Van der Pol oscillator.  The new, third degree of freedom corresponds to the atomic inversion, which is no longer constant at large detunings or drive strengths.  We show qood quantitative understanding of our system, providing a solid foundation for future work in fundamental physics using superradiant lasers. 

The apparatus for our superradiant laser based on Raman transitions has been described in previous work \cite{Bohnet2012,Bohnet2012a,Weiner2012,Bohnet2013}.  A conceptually simplified experimental diagram for this work is shown in Fig.~\ref{fig:fig1}.  

\begin{figure}[h!]
\includegraphics{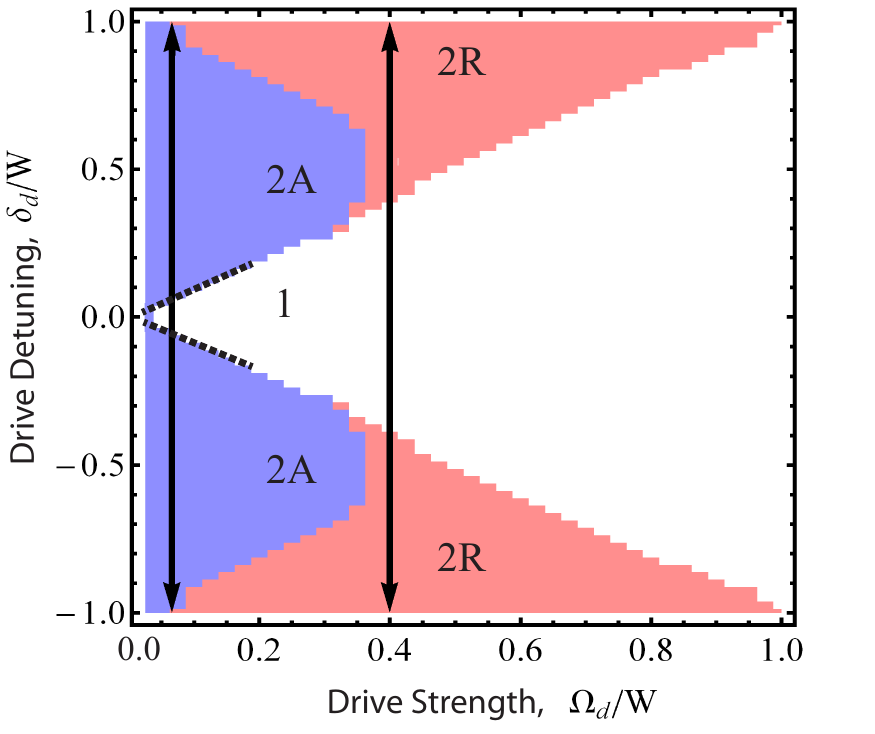}
\caption{\label{fig:fig2} The predicted phase diagram for the driven superradiant laser is shown in a plane defined by the applied drive strength $\Omega_d$ and the drive detuning $\delta_d$, normalized to the repumping rate $W$, which is fixed to $N C \gamma/2$ here.  The regions are first divided by the number of distinct emission frequencies (1 or 2). Region (2) is further divided by the frequency shift of the self-lasing component at $\omega_\ell$, which can be attracted (2A) or repelled (2R) from the applied drive frequency $\omega_d$.  When $\Omega_d<0.2\times W$, the laser synchronizes by smoothly coalescing in frequency with the drive (dashed line).  For larger drives, the self-lasing component remains distinct and is quenched.  The two trajectories (black arrows) refer to the two parameter trajectories explored by the data in Fig.~\ref{fig:fig3}.}
\end{figure}

The key distinct feature in this work is the application of an additional coherent drive to the superradiant laser's cavity mode (Fig.~\ref{fig:fig1}).  The drive couples the upper $\ket{\uparrow}$ and lower $\ket{\downarrow}$ lasing states with a single-atom Rabi frequency $\Omega_d$.  The drive frequency $\omega_d$ is detuned from the effective atomic transition frequency $\omega_a$ by $\delta_d\equiv \omega_d - \omega_a$ (Fig.~\ref{fig:fig1}(b)).  The behavior of the system depends on the relative magnitudes of drive strength $\Omega_d$, detuning $\delta_d$, and characteristic rates of the superradiant laser: the repumping rate $W$ and characteristic collective emission rate into the cavity $N C \gamma$ given by the collective cooperativity $N C$ and the single-atom decay rate $\gamma$ from $\ket{\uparrow}$ to $\ket{\downarrow}$  (see Supplementary Material for details).

When no drive is applied the laser emits at frequency $\omega_{\ell_0}$ near, but not necessarily identical to $\omega_a$. When the drive is applied, the lasing frequency $\omega_{\ell}$ is shifted by the atoms' interaction with the drive by $\delta_\ell\equiv\omega_{\ell} -\omega_{\ell_0}$. Additionally, the laser can emit at the drive frequency $\omega_d$.  We detect the light emitted from the cavity using heterodyne detection.  This gives complete information about the emission spectrum and allows us to measure $\omega_{\ell}$, $\omega_d$ and the phases and amplitudes of the electric fields emitted from the cavity at these frequencies. Other frequency components in the laser emission are expected and observed at sums and differences of $\omega_{\ell}$ and $\omega_d$.  These additional components can become rather large near synchronization (see Ref. \cite{doi:10.1142/S0218127400001481}), but detailed treatment is beyond the scope of this work.

The predicted behavior of the emitted field of the laser is summarized by the theoretical phase diagram in Fig.~\ref{fig:fig2}.  The phase diagram is calculated by numerically integrating optical Bloch equations based on a simplified 2-level model for the superradiant laser (see \cite{Bohnet2014} and Supplementary Material for details).  For simplification, the repumping rate $W$ is set to a value that optimizes the output power of the laser $W_{\text{opt}} = \frac{1}{2} N C \gamma$ \cite{Meiser2009}, so that the characteristic rates governing the phase diagram are the two ratios $\Omega_d/W$ and $\delta_d/W$.  

\begin{figure}[h!]
\includegraphics{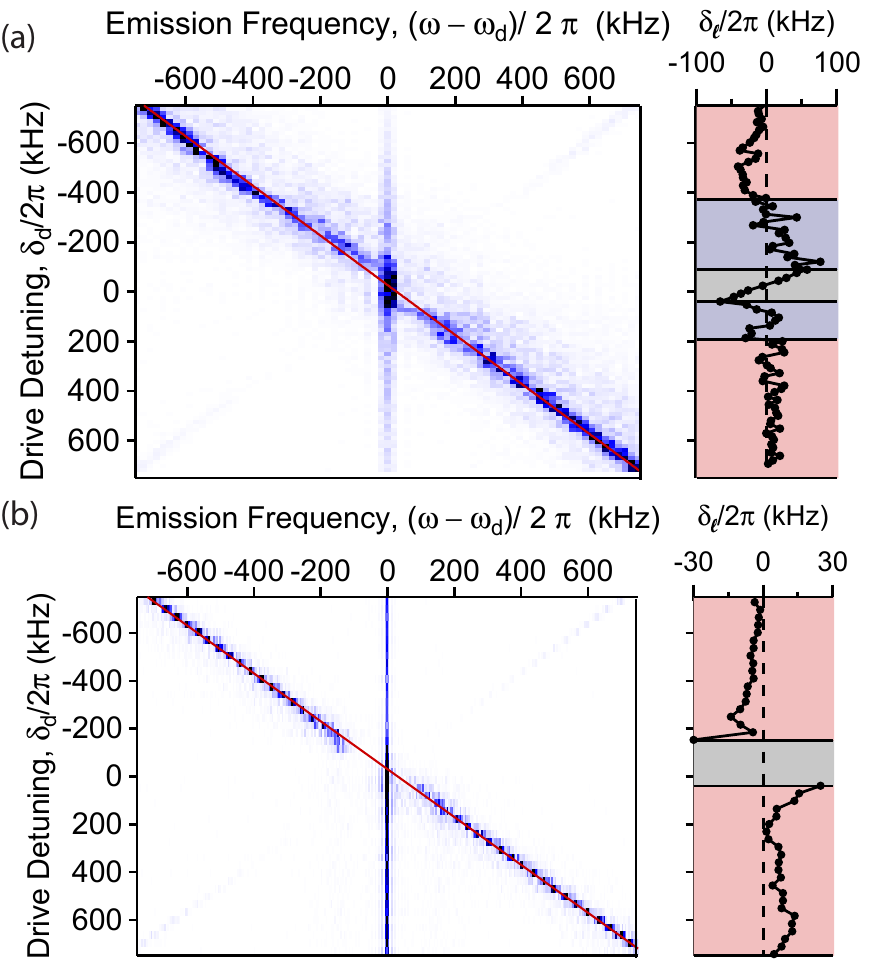}
\caption{Experimental observation of coalescing attractive (a) and repulsive (b) synchronizations. (Left) 2D spectrograms are taken with fixed drive strength $\Omega_d$ as the detuning of the drive $\delta_d$ is varied along the representative vertical trajectories in Fig.~\ref{fig:fig2}. Darker colors indicate higher power in a frequency bin.  The red line indicates the expected self-lasing trajectory in the absence of an applied drive.  (Right) Two panels show the frequency shift $\delta_\ell= \omega_\ell-\omega_{\ell_0}$ between the lasing frequency and the lasing frequency when no drive is present.  In these plots, each region, attraction and repulsion is colored similarly to the phase diagram Fig.~\ref{fig:fig2} and qualitatively follows the predicted behaviors for the two distinct trajectories.}
\label{fig:fig3}
\end{figure}

The primary feature of the phase diagram is the synchronization or non-synchronization of the superradiant emission to the drive.  In the unsynchronized phase of region (2), the atomic dipoles are not perfectly synchronized to the drive and the spectrum of light contains two distinct frequency components at the drive frequency $\omega_d$ and the self-lasing frequency $\omega_{\ell}$. In contrast in region (1), the atomic dipoles become synchronized to the drive and all light emission occurs at $\omega_d$.  For the optimum repumping $W=W_{\text{opt}}$ here, the synchronization transition occurs roughly when $\Omega_d = \delta_d$. 

The unsynchronized region (2) is broken into two subregions, delineated by the self-lasing's attraction toward or repulsion from the drive frequency. The region of attraction ($\frac{\delta_\ell}{\delta_d}>0$) is labeled (2A) in the phase diagram.  The region of repulsion ($\frac{\delta_\ell}{\delta_d}<0$) is labeled (2R).  As $\delta_d$ or $\Omega_d$ are tuned, the behavior of the approach to synchronization  depends on whether one enters region (1) from region (2A) or (2R).  For attractive synchronization (from (2A) to (1)), as the $\Omega_d = \delta_d$ boundary is crossed, $\omega_{\ell}$ is pulled toward the drive. For drive strengths  $\Omega_d < 0.2 \times W$ (dashed line in Fig \ref{fig:fig2}),  the self-lasing component synchronizes by smoothly coalescing with the drive at $\omega_d$.  When $\Omega_d> 0.2\times W$, the self-lasing is driven to zero before coalescence can occur.  In the repulsive synchronization (from (2R) to (1) in the phase diagram), as one approaches synchronization $\omega_{\ell}$ is repelled in frequency from $\omega_d$ and the self-lasing component is driven to zero so that the superradiant ensemble is emitting power only at the drive frequency $\omega_d$.  

In the limit $\Omega_{d}, |\delta_d| \ll W$, the laser inversion $J_z$ is approximately fixed, and the three-dimensional Bloch vector ($J_x$, $J_y$, $J_z$) describing the atomic ensemble can be reduced to an effective two dimensional object described by the transverse coherence $J_-= J_x - i J_y$ (see Supplementary Material for more detail and a Bloch vector interpretation of the phase diagram).  In this case, the equation describing the time evolution of $J_-$ is closely equivalent to that of a driven Van der Pol oscillator, for which attractive synchronization ((2A) to (1)) with and without coalescence (characterized by saddle-node and Hopf bifurcations respectively) have been well studied \cite{doi:10.1142/S0218127400001481,pikovsky2003synchronization}.  

However, when either $\Omega_d$ or $|\delta_d|\gtrsim W$, the inversion $J_z$ can no longer be approximated as fixed and the dynamic response of the full three-dimensional Bloch vector must be considered. The response of the extra degree of freedom $J_z$ leads to the repulsive behavior in region (2R) which can be interpreted as an AC Stark shift (see Supplementary Material for more explanation).

Experimental examples of the two synchronization transitions are shown in Fig.~\ref{fig:fig3}, with approximate  trajectories in the phase diagram represented by black arrows in Fig.~\ref{fig:fig2}.  Fig.~\ref{fig:fig3}(a) demonstrates attractive, coalescing synchronization and Fig.~\ref{fig:fig3}(b) represents repulsive synchronization.  The left plots are two-dimensional power spectra of the laser emission. Each horizontal slice corresponds to a single power spectrum of laser emission where color represents the optical power in each frequency bin.  The nominal detuning $\delta_d$ is changed between experimental trials and plotted on the left axis.  On the horizontal axis, the drive freqency $\omega_d$ is set to zero so that in the absence of an applied drive ($\Omega_d =0$), the emission frequency $\omega_\ell$ would follow the diagonal red line.  At sufficiently small detunings synchronization occurs and only power at $\omega_{d}$ is observed.

The spectrograms illustrate the qualitative differences between the two types of synchronization. In Fig.~\ref{fig:fig3}(a), the self-lasing frequency $\omega_\ell$ is attracted toward and joins $\omega_d$ as $\left|\delta_d\right|$ becomes small.  In Fig.~\ref{fig:fig3}(b), the two emission components remain distinct until the $\omega_\ell$ component is extinguished.  To more clearly illustrate the attraction and repulsion, the measured quantity $\delta_\ell = \omega_{\ell}-\omega_{\ell_0}$ is plotted on the right. The plots are overlaid with color to help identify the  repulsion (red) and attraction regions (blue) matching the phase diagram of Fig.~\ref{fig:fig2}.  The sizes of each region of behavior are not expected to precisely match the simplified phase diagram for a 2-level laser due to additional dispersive cavity tuning effects present in the Raman system, but the phase diagram captures the qualitative behavior of the data.  For the trajectory at small $\Omega_d$ the laser goes from synchronized (1), to attraction (2A), to repulsion (2R) as predicted.  For the trajectory at larger $\omega_d$ there is a single transition from (1)- to (2R)-like responses.  

\begin{figure}[h!]
\includegraphics{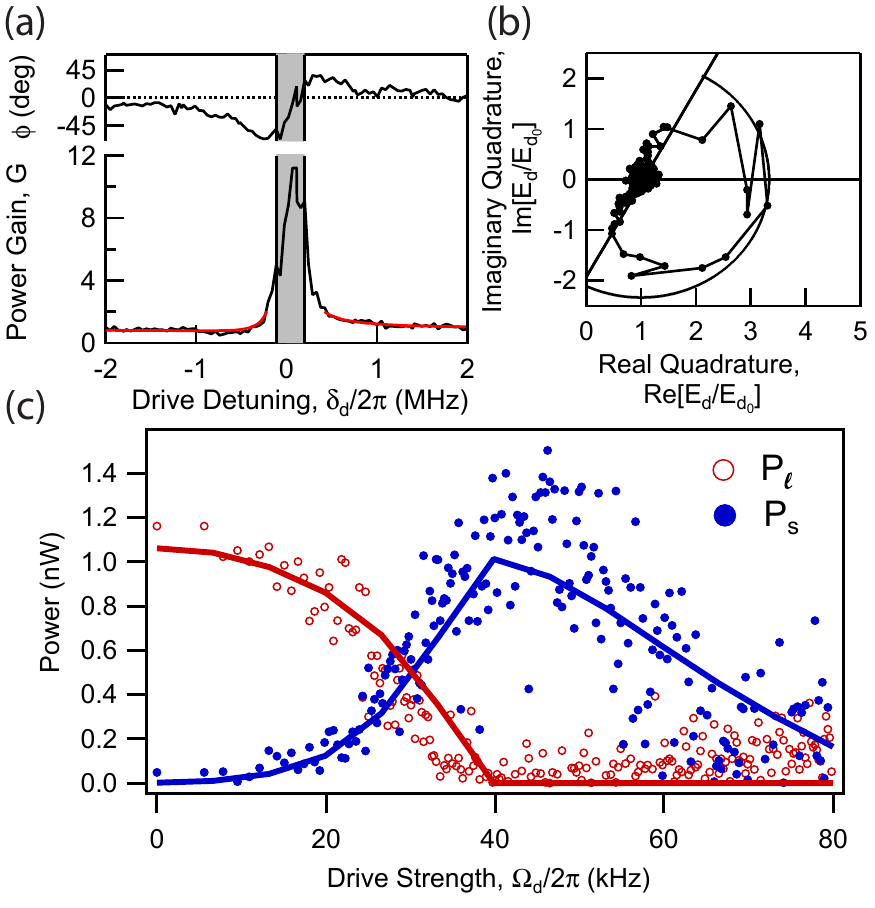}
\caption{\label{fig:fig4} Synchronization and gain saturation. (a) Measured gain and phase response of the superradiant laser at the drive frequency $\omega_d$.  At large drive detunings, the response displays linear small-signal gain (red fit to perturbative model overlayed).  The gain saturates (grey shaded region) at small detunings as the laser approaches the synchronization transition.  (b) The same gain and phase response are represented in a phasor picture.  Points of small signal gain lie along the straight line, and the region of saturation is approximately described by a curve of maximum stimulated electric field (inner curve).  (c)  The stimulated output powers $P_s$ and $P_\ell$, at the self-lasing frequency $\omega_\ell$ (red) and at the drive frequency $\omega_d$ (blue) respectively, are displayed as the laser is driven across a repulsive synchronization transition at approximately $\Omega_d / 2 \pi \approx 40$~kHz.  Theoretical predictions (solid lines) show good agreement with the data.}
\end{figure}

We now turn to the development of a perturbative description of the system far from synchronization and the break down of this description as the system approaches and ultimately crosses the synchronization threshold.

Deep into region (2) of the phase diagram, the superradiant laser's response to the drive is small and can be understood as a small modulation of the initially undriven Bloch vector describing the atomic coherence.  The modulated Bloch vector then radiates an additional field into the cavity at the drive frequency $\omega_d$, producing gain. In Fig.~\ref{fig:fig4}(a) we measure this power gain $G$ and phase response $\phi$ of the laser at the drive frequency $\omega_d$ versus the drive detuning $\delta_d$.  This corresponds to a vertical trajectory on the phase diagram where, in this dataset, $\frac{\Omega_d}{W} \approx 0.04$.  A fit to a perturbative model based on the optical Bloch equations is shown in red (see Supplementary Material).

At small detunings (grey region), the gain and phase response begin to saturate, and deviate from the predicted small-signal values.  This is roughly the point when the laser begins to synchronize to the drive and the emission at frequencies other than $\omega_d$ begin to disappear.

In Fig. \ref{fig:fig4}(b), the equivalent complex electric field response $E_d$ at the drive frequency is shown in a phasor diagram with each point corresponding to the measured field at a given detuning in Fig. \ref{fig:fig4}(a).  The drive response when no atoms are present $E_{d_0}$ is normalized to be real and of length $1$.  In the perturbative limit, the additional stimulated field follows a straight line.  The line of small-signal gain is tilted due to an additional phase shift of the stimulated field arising from nonzero detuning of the drive from the optical cavity $\omega_d-\omega_c \neq 0$.  At saturation, the stimulated field deviates from the straight line and qualitatively follows a contour of constant stimulated electric field (solid semi-circle).

In Fig.~\ref{fig:fig4}(c) we show an example of how optical power is ``stolen" from the self-lasing frequency $\omega_\ell$ and transferred to the drive frequency.  Here the drive strength $\Omega_d$  is increased at a fixed detuning $\delta_d/W= 2.2$, and the vertical axis shows the self-lasing power $P_\ell$ and the stimulated drive power $P_s\equiv P_d - P_{d_0}$.   $P_{d_0}$ is the detected power at the drive frequency in the absence of any atomic response, scaling as $P_{d_0}\propto \Omega_d^2$, such that $P_s$ represents the extra stimulated power at $\omega_d$.   This dataset corresponds to tuning the system along a horizontal line in Fig.~\ref{fig:fig2} that lies outside the plotted range and such that the system crosses from the repulsive region (2R) to the synchronized region (1).

Numerical solutions (solid lines in Figure~\ref{fig:fig4}(c)) of the optical Bloch equations (Supplementary Material) give reasonable agreement with the data.  The theoretical model includes approximate corrections for an additional cavity tuning effect \cite{Bohnet2013}, and the absolute vertical scale of the theory has been scaled so that the $P_\ell$ agrees with the data at $\Omega_d = 0$.

The synchronization point in this data is represented by the sharp point when $P_\ell$ hits zero, with a discontinuous first derivative in $P_\ell$ and $P_s$.  At the synchronization point, $P_s$ is approximately the original output power of the laser when $\Omega_d$ is zero.  At large drive strengths in Fig.~\ref{fig:fig4}(c), $\Omega_d$ becomes much larger than $W$ and the total output power of the laser decreases due to repumping-induced dephasing of the rapid Rabi oscillations caused by the drive..

We have observed for the first time two different types of synchronization transitions of a superradiant laser to an external drive, one attractive and one repulsive in nature. The synchronization transition is analogous to a ferromagnet in the presence of an applied magnetic field, the drive breaking a continuous symmetry of the laser with respect to phase \cite{DeGiorgio1970}.  However, the laser steady state is far from thermodynamic equilibrium, making our well-controlled cold atom-cavity system an interesting avenue for continued study of nonequilibrium phase transitions with modern approaches \cite{PhysRevA.87.023831}.

It is often useful to apply an external drive to a superradiant laser.  Such drives have been used, for instance, to  probe the frequency of the optical cavity  in Raman-laser systems such as ours (as was done in ref. \cite{Bohnet2012a}) or perhaps in future narrow linewidth superradiant lasers to reduce errors, inaccuracies, and technical noise due to cavity frequency pulling.  This work establishes understanding for how such a technical probe will affect the system.  Furthermore, the phase response within the saturation region of Fig.~\ref{fig:fig4}(a)  could be used as an error signal for a form of active spectroscopy of the gain medium, in some sense, the inverse approach to that of Ref. \cite{nonlinear}, although the fundamental signal-to-noise of such an approach is an open question.  

In the future, the understanding provided in this work will guide the interpretation of other proposed experiments in cold atom-cavity systems.  For instance, multiple superradiant sub-ensembles each with an independent transition frequency $\omega_a$ can be engineered to interact with each other through one or multiple cavity modes \cite{PhysRevLett.107.277201,PhysRevLett.107.277202,doi:10.1080/14786435.2011.637980,Xu2014,Manzano2013,PhysRevLett.112.094102}.   Furthermore, while our phase diagram and the average response of a superradiant laser to an external drive are well-described by a mean field description, and therefore can be considered classical behavior, recent theoretical works propose systems of multiple superradiant ensembles where quantum noise becomes observably large and may serve to drive the phase transitions and affect the average behavior \cite{Xu2014,PhysRevLett.112.094102,Manzano2013}.  
\begin{acknowledgments}
The authors would like to acknowledge helpful discussions with Matthew Norcia, Justin Bohnet, Daniel Gauthier, Ana Maria Rey, and Murray Holland.  All authors acknowledge financial support from DARPA QuASAR, ARO, NSF PFC,
and NIST. K.C.C. acknowledges support from NDSEG. This work is supported by the National Science Foundation under Grant Number 1125844.
\end{acknowledgments}

\bibliography{main}

\pagebreak
\widetext
\begin{center}
\textbf{\large Supplemental Material: Phase diagram for injection locking a superradiant laser}
\end{center}
\setcounter{equation}{0}
\setcounter{figure}{0}
\setcounter{table}{0}
\setcounter{page}{1}
\makeatletter
\renewcommand{\thefigure}{S\arabic{figure}}

\section{Experimental Setup}
The apparatus for the superradiant laser and principles behind its basic operation have been described in detail in previous work \cite{Bohnet2012,Bohnet2012a,Weiner2012,Bohnet2013}.  The atomic gain medium consists of $N\approx1.1\times 10^6~^{87}$Rb atoms cooled to 10 to 20~$\mu$K and trapped in a 1D optical lattice inside of an optical cavity with power decay linewidth $\kappa =2\pi\times11.8$~MHz.  The atoms are tightly confined to $\ll \lambda$ (i.e. the Lamb-Dicke regime) along the cavity axis, but only weakly confined perpendicular to the cavity axis.

A dressing laser is applied transverse to the cavity to induce spontaneous Raman transitions between two hyperfine gound states $\ket{\uparrow}\equiv\ket{5S_{1/2},F = 2,m_F = 0}$ to $\ket{\downarrow}\equiv\ket{5S_{1/2},F = 1, m_F = 0}$, with typical single-atom free-space Raman transition rates $\gamma=2 \pi \times 100~$Hz to $2 \pi \times 300~$Hz.  The cavity frequency $\omega_c$ is tuned to be on or near resonance with the spontaneously emitted light's frequency $\omega_a$. The effective two-photon coupling to the cavity is characterized by the rms value of the Jaynes-Cummings coupling constant $g_2$ \cite{Meiser2010} and single atom cooperativity parameter $C = \frac{4 g_2^2}{\kappa\gamma} = 5\times10^{-3}$.  The collective (or superradiant) emission rate for a single atom scales as $NC\gamma$.




To maintain population inversion and steady-state emission, additional lasers are applied to incoherently repump atoms through optically excited states from $\ket{\downarrow}$ back to $\ket{\uparrow}$.  The characteristic repumping rate from $\ket{\downarrow}$ (including Rayleigh scattering) is $W\approx 2\pi\times60~$kHz to $2\pi\times500~$kHz. The repumping process is the primary contribution to the atomic transverse decoherence rate $\gamma_\perp \approx W/2 + \Gamma_D$.  We measure a small additional contribution to the transverse broadening $\Gamma_D$ which is primarily due to doppler broadening of the two-photon transition from $\ket{\uparrow}$ to $\ket{\downarrow}$ resulting from the weak transverse confinement of the atoms.

\section{Optical Bloch Equations} \label{Gain}

The average behavior of the superradiant laser with an applied drive can be understood with slight modifications to the optical Bloch equations presented in \cite{Bohnet2014}.  The optical Bloch equations describe the time evolution of expectation values of the cavity annihilation operator $\hat{a}$ and the collective atomic operators, $\hat{J}_z$ and $\hat{J}_-$, defined as,
\begin{align}
\hat{J}_z &= \sum_{\substack{i = 1}}^N \frac{\ket{\uparrow_i}\bra{\uparrow_i}-\ket{\downarrow_i}\bra{\downarrow_i}}{2}\\
\hat{J}_- &=\sum_{\substack{i = 1}}^N \ket{\downarrow_i}\bra{\uparrow_i}.
\end{align}
$\hat{J}_z$ and $\hat{J}_-$ represent the atomic inversion and transverse coherence respectively.  The optical Bloch equations govern the time evolution of the expectation values of these operators, $J_z = \langle\hat{J}_z\rangle$, $J_- = \langle\hat{J}_-\rangle$, and $E = \langle\hat{a}\rangle$.  The atomic response can be visualized as a three-dimensional Bloch vector with  x and y projections of the vector given by $J_-= (J_x-iJ_y)/2$.  $E$ is a complex representation of the optical cavity electric field such that $|E|^2$ is the average number of photons inside the cavity.  The nonlinear equations are closed by approximating that  the expectation values of products of operators can be factorized into products of expectation values.  Assuming uniform coupling to the cavity mode, the coupled equations for a 2-level system with an applied drive can be written as

\begin{align}
\label{bloch}
\dot{E} &= -\left(\frac{\kappa}{2}+i \delta_c\right)E - i g_2 J_-+\frac{\kappa}{2} E_{d_i}\; e^{i \delta_d t}\\
\dot{J}_- &= - \gamma_\perp J_- + i 2 g_2  J_z E\\
\dot{J}_z &= -W J_z + \frac{N}{2} W + i g_2 (J_- E^* - J_-^* E).
\end{align}

The equations are written in a frame rotating at the atom's natural transition frequency $\omega_a$.  
$E_{d_i}$ is proportional to the amplitude of the electric field of the applied drive incident on the optical cavity.  $\delta_c = \omega_c-\omega_a$ is the detuning of the cavity from the natural atomic transition frequency $\omega_a$.  The rest of the equation parameters are defined in previous text.  The Rabi frequency of the drive is related to these parameters by $\Omega_d = \frac{2 g_2 E_{d_i}}{1+\tilde{\delta}_c^2}$, where quantities  $\tilde{A} \equiv \frac{A}{\kappa/2}$. The measured intracavity field when no atoms are present is $E_{d_0} = \frac{E_{d_i}}{1+i \tilde{\delta}_{d,c}}$, where $\delta_{d,c}$ is the detuning of the driving electric field from the cavity resonance, $\delta_{d,c} = \omega_d-\omega_c$.

We have numerically solved these equations to derive the phase diagram of Fig.~2 and to create theoretical curves for Fig.~4.  The two-level model does a reasonable job of predicting the  behavior of synchronization versus $\Omega_d$ and $\delta_d$.  However, due to the true multi-level structure of the laser, these equations cannot be used to predict the total output power of the system.  When these equations are not adequate, multi-level optical Bloch equations from Ref. \cite{Bohnet2014} can be used.  Additionally, to approximately account for dispersive shifts of the optical cavity in our Raman laser, the cavity detuning $\delta_c$ is made a function of $J_z$ to generate the theory for Fig. 4(c).  Details can be found in Ref. \cite{Bohnet2014}.

\section{Small-signal gain}

The small-signal regions of gain and phase response, as described in the main text, follow a simple form that can be derived from the optical Bloch equations.  We first assume $\kappa$ is larger than all other characteristic frequencies in the system so that $E$ adiabatically follows $J_-$.  We then assume that $J_-$ primarily responds at two frequencies, $\omega_a$ and $\omega_d$, and make the ansatz $J_- = J_{-a}+J_{-d}~e^{i \delta_d t}$.   Lastly, we assume that in this perturbative limit $J_z$ is unaffected by the weak drive, and retains its steady state value with no drive, $J_{z_{ss}} = \frac{W (1+ \tilde{\delta}_c^2)}{2 C \gamma }$.  With these approximations, we find the small-signal complex field response at the drive frequency.  The total detected field at the drive frequency, $E_d= E_{d_0}+E_s$, has two contributions.  One contribution ($E_{d_0}$) comes from the drive alone, and the other ($E_s$) from atomic stimulation.  The small-signal form of $E_s$ is given by

\begin{align}
\label{as}
E_s &=\frac{- i  E_{d_0}\gamma_\perp(1+\tilde{\delta}_c^2)}{\delta_d(1+i \tilde{\delta}_c)}.
\end{align}
In the limit $\tilde{\delta}_c = 0$, $E_s$ can be written in the form $E_s = \frac{-i \sqrt{\alpha} E_{d_0}}{\delta_d}$, where $\sqrt{\alpha}$ characterizes the stimulated field gain.  From Eq. \ref{as} one expects $\alpha = \gamma_\perp^2$.  

Eq. \ref{as} explains several features of the measured small-signal gain shown in Fig. 4 of the main text.  For $\delta_c=0$, the stimulated field is in the orthogonal quadrature to the driving field.  However, when $\delta_c\neq 0$ as is the case in the data of Fig.~4(b), the response of the cavity to the driving atomic dipole causes the stimulated field to be partially rotated into the same quadrature as the driving field.  This rotation breaks the symmetry about $\delta_d=0$  for both the measured phase and power gain shown in Fig. 4(a).


\begin{figure}
\centerline{\includegraphics{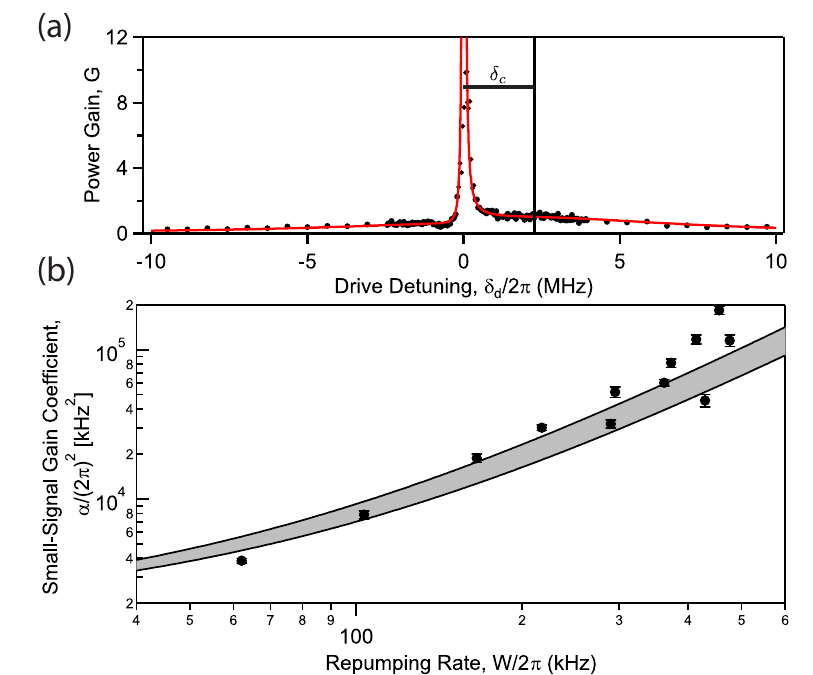}}
\caption{\label{fig1}Quantitative small-signal gain measurements.  (a) We measure the total transmitted power at the drive frequency $\omega_d$ and define power gain $G$ as the measured transmitted power normalized to the drive power transmitted through the cavity on resonance with no atoms present.  For these measurements $\delta_d$ is scanned over a frequency range greater than the cavity linewidth $\kappa$.  The data is fit to the model in Eq. \ref{GainFit} (red line).  Note that here, the cavity resonance marked by the vertical solid line is a few MHz higher than the atomic resonance.  (b) From fits to data such as (a), we plot the variation of the fitted gain coefficient $\alpha$ (see text) versus $W$.  The prediction $\alpha = \gamma_\perp^2$ is shown in grey.  The width of the grey band corresponds to the uncertainty in an independant calibration of the repumping rate $W$.}
\end{figure}

To quantify the small-signal gain experimentally, we measure the total output power of the laser at the drive frequency for a large range of drive detunings $\delta_d$.  We fit the total output power to the gain model shown in Eq. \ref{GainFit}.  Equation \ref{GainFit} is the magnitude squared of Eq. \ref{as} with the $E_{d_0}$ dependence on $\tilde{\delta}_{d,c}$ written explicitly.


\begin{align}
\label{GainFit}
G(\delta_d, \tilde{\delta}_{d,c}) &= \frac{G_0}{1+\tilde{\delta}_{d,c}^2}~\Bigg\{ 1-2 \frac{\sqrt{\alpha} (\tilde{\delta}_c-\tilde{\delta}_0)}{(\delta_d-\delta_0)}
+\frac{\alpha[1+(\tilde{\delta}_c-\tilde{\delta}_0)^{2}]}{(\delta_d-\delta_0)^2}\Bigg\}
\end{align}
For a single scan of $\delta_d$, we allow fitting of the parameters $G_0$, $\alpha$, $\delta_c$, $\kappa$, and $\delta_0$.  The fit model constrains the transmitted power to be $G_0$ when the drive is on resonance with the cavity in the absence of an atomic response.  The $\delta_0$ coefficient allows for an arbitrary offset of the atomic transition frequency $\omega_a$ from zero.  Figure~\ref{fig1}(a) displays an example of this measured total output power as a function of $\delta_d$ with the fit overlaid in red.  In Fig.~\ref{fig1}(a), the data has been rescaled such that the fitted coefficient $G_0 = 1$.  After this rescaling, the data represents the total power emitted at the drive frequency, normalized to the power transmitted through the cavity when the drive is on resonance with the cavity and no atoms are present.   Also, the frequency axis of the data has been adjusted such that $\delta_0 =0$.


We follow this procedure, measuring the output power and fitting to Eq.~\ref{GainFit}, for many repumping rates $W$.  We plot the fitted gain coefficients $\alpha$ versus $W$ in Fig.~\ref{fig1}(b).  The prediction that $\alpha = \gamma_\perp^2 = (W/2+\Gamma_D)^2$ is overlaid in grey.    Uncertainty in the prediction (width of the grey band) is due to uncertainty in the experimental calibration of $W$.  The prediction shows reasonable agreement with the theory over a significant range of $W$.

\section{Bloch Vector Interpretation of Phase Diagram}
\begin{figure}
\includegraphics{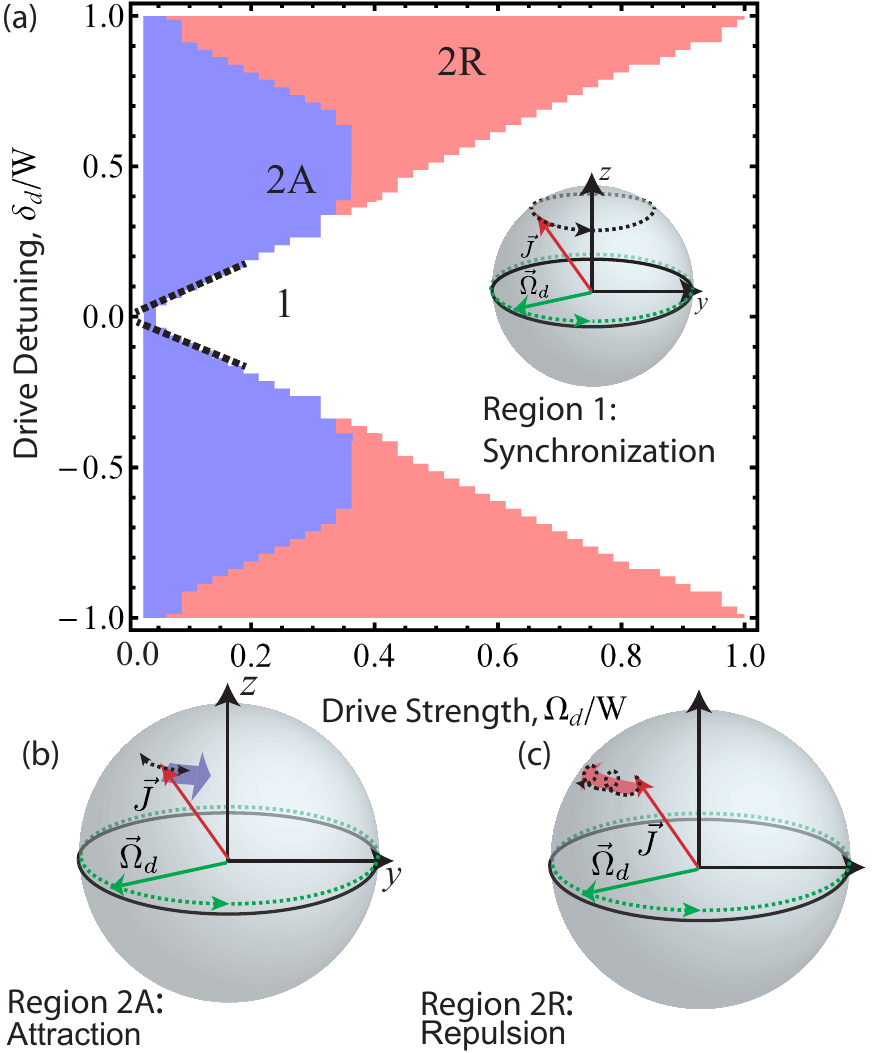}
\caption{\label{fig2} Bloch vector interpretation of phase diagram. (a) The types of behavior for the driven superradiant laser can be characterized by a phase diagram.  The characteristic rates that determine the lasing behavior are drive Rabi frequency $\Omega_d$, detuning $\delta_d$, and repumping rate $W$.  The regions correspond to the number of distinct emission frequencies (1 or 2) and the frequency shift (attraction or repulsion) of the carrier (A and R respectively).  The behavior of the synchronization (a), attraction (b), and repulsion (c) configurations are shown in a Bloch sphere picture.  In the frame of the atomic transition frequency $\omega_a$, the drive is represented by a rotation $\vec{\Omega}_d$, with an orientation which rotates along the dashed green trajectory at frequency $\delta_d$.  In the unsynchronized case this modulates the Bloch vector (red vector), causing drift toward or away from the drive, with the average precession of the Bloch vector indicated in each case via the large blue and red arrows.  In the synchronized case, the Bloch vector follows the drive all the way around the sphere.}

\end{figure}

The predicted phase diagram is shown in Fig. S2.  The behavior in each of the 3 regions can be visualized by the behavior of the Bloch vector in each regime.  In a frame rotating at the atom's natural transition frequency $\omega_a$, the applied drive can be represented by a rotation of the Bloch vector, $\vec{\Omega}_d = \Omega_d\left(\hat{\mathbf{x}}\cos(\delta_d t) + \hat{\mathbf{y}}\sin(\delta_d t)\right)$.  $|\Omega_d|$ is the angular frequency of the rotation, and $\hat{\Omega}_d$ is the axis about which the Bloch vector rotates.  The azimuthal phase of the applied rotation axis precesses at frequency $\delta_d$.  When $\Omega_d\ll\left|\delta_d\right|$, the drive primarily acts to slightly modulate the orientation of the Bloch vector (both Fig.~\ref{fig2}(b) and (c)).  However, when $\Omega_d>\left|\delta_d\right|$, the applied modulation is so large that the Bloch vector can actually follow the drive all the way around the sphere.  This is the synchronized region (1) in Fig. \ref{fig2}(a).  Near the synchronization transition,  the repulsive (2R) versus attractive (2A) behavior is determined by the size of the repumping rate $W$ compared to $\delta_d$ and $\Omega_d$. 

\subsection{Attractive Regime: Mapping to 2D Van der Pol oscillator}
In the case $|\delta_d|,\Omega_d\lesssim W$ (i.e. 2A) the drive does not significantly perturb the laser from its steady-state inversion because any change in $J_z$ caused by the applied field is quicky healed by the repumping process \cite{Bohnet2012a}. When $J_z$ is not modified, the azimuthal phase $\phi$ is partially or fully dragged in the same direction as the rotating axis $\hat{\Omega}_d$ (Fig.~\ref{fig2}(a)).  The lasing can be viewed as being captured by the applied drive.

Furthermore, in this regime of weak drive, $|\delta_d|,\Omega_d\lesssim W$, the system can be mapped onto a Van der Pol self-oscillator model with a nonlinear driving term,
 \begin{align}
 \label{vdp}
 \dot{\jmath}_-=-i \delta \jmath_- + \lambda \jmath_-  (1-|\jmath_-|^2)+\Omega(1-\beta |\jmath_-|^2)
 \end{align}
with complex amplitude $\jmath$ and characteristic rates $\lambda$, $\delta$, $\Omega$, and $\beta$.  Equation \ref{vdp} has an equilibrium amplitude $|\jmath_-|^2=1$.  The nonlinearity of the applied drive is governed by the parameter $\beta$.  For $\beta = 0$ this model has been well studied \cite{doi:10.1142/S0218127400001481,pikovsky2003synchronization}.

To explicitly show the mapping of the optical Bloch equations (Eq.~\ref{bloch}) onto this form, we first eliminate the cavity field $E$ by assuming operation in the deep bad cavity limit, and assume that $J_z$ is not perturbed by the drive.  Setting $\dot{J_z}$ equal to zero gives the nominal steady state value, 
\begin{align}
J_{z_{ss}} = \frac{N}{2}-\frac{C \gamma}{W} |J_-|^2.
\end{align}
One can then insert $J_{z_{ss}}$ into the $J_-$ equation in Eq. \ref{bloch}, which leads to
\begin{align}
\label{mappingeq}
\dot{J}_- &= -i \delta_d J_-  +J_-\left[\left(\frac{N C \gamma}{2}- \frac{W}{2}\right)-\frac{(C\gamma)^2}{W} |J_-|^2\right]  \\
&+ \frac{\Omega_d}{2}\left(\frac{N}{2} - \frac{C \gamma}{W} |J_-|^2\right)\notag.
\end{align}
Eq. 10 is identical in form to Eq. \ref{vdp}.  For the case of optimal repumping, we set $W = W_{\text{opt}} = \frac{N C \gamma}{2}$, and normalize $J_-$ to its steady state value defining $\jmath_- = J_-/J_{-_{ss}}$, where $J_{-_{ss}}= \frac{N}{\sqrt{8}}$, giving

\begin{align}
\dot{\jmath}_- &= -i \delta_d \jmath_- + \frac{N C \gamma}{4} (1-|\jmath_-|^2) + \frac{\Omega_d}{\sqrt{2}}\left(1-\frac{|\jmath_-|^2}{2}\right).
\end{align}
This equation is of the same form as Eq.~\ref{vdp} with $\delta = \delta_d$, $\Omega = \frac{\Omega_d}{\sqrt{2}}$, $\beta =\frac{1}{2}$, and $\lambda =\frac{NC\gamma}{4}$.  We can define an effective drive strength $\Omega_d'=\frac{\Omega_d}{\sqrt{2}}\left(1-\frac{|\jmath_-|^2}{2}\right)$.  For a weak drive, the laser remains close to its steady state.  One finds $\Omega_d'=\frac{1}{\sqrt{8}} \Omega_d$, and the system can be thought of as behaving similarly to the standard driven Van der Pol oscillator of Ref.~\cite{doi:10.1142/S0218127400001481,pikovsky2003synchronization} with a constant driving term.
\subsection{Repulsive Regime: 3D dynamics}
When the drive is applied with a large detuning $|\delta_d|\gtrsim W$ (i.e. (2R)), the repumping at rate $W$ cannot heal the changes in the inversion $J_z$ caused by the applied drive \cite{Bohnet2012a}.  $J_z$ can no longer be considered static and thus introduces a third degree of freedom (in addition to $J_x$ and $J_y$) in the system.  In this regime, the Van der Pol model breaks down, and small oscillations in $J_z$ must be taken into account.  The unhealed modulations of $J_z$ at frequency $\delta_d$ coherently interact with the applied rotation due to the drive to cause the Bloch vector to on average aquire a small precession in the opposite sense to the precession of the drive rotation axis $\hat{\Omega}_d$ (shown in Fig. \ref{fig2}(c)).  The precession rate is second order in $\Omega_d$ and can be identified as an AC Stark shift that leads to the observed frequency repulsion in region (2R).  To emphasize, this shift does not appear in region (2A) because there the repumping process acts to smooth out the  modulations in $J_z$ that are essential for creating the AC Stark shift.

The AC Stark shift can be derived to leading order by perturbatively allowing for small oscillations in $J_z$.  In this way we can mathematically show the additional repulsive behavior not evident in the 2-dimensional model of Eq. \ref{vdp}.  The optical Bloch equation for $J_z$ allowing for modulation, and written in the frame of the drive frequency $\omega_d$ is,
\begin{align}
\dot{J_z}=  -W J_z +\frac{N}{2} W -C\gamma |J_-|^2- \Omega_d \text{Re}(J_- )
\end{align}
We treat this equation perturbatively by assuming that $J_-$ primarily oscillates at the self-lasing frequency.  From this we can derive a leading order repulsion term giving a new self-lasing frequency,

\begin{equation}
\delta' = \delta_d+\frac{\Omega_d^2 \delta_d }{2(\delta^2_d+W^2)},
\end{equation}
which arises from oscillations in $J_z$ coupling into an average frequency repulsion of the Bloch vector.  This repulsive physics is not present in the driven Van der Pol oscillator and arises from the Bloch vector occupying a higher, 3-dimensional parameter space.

\bibliography{main}

\end{document}